\documentclass[prd,showpacs,letterpaper,twocolumn]{revtex4}%
\usepackage{graphicx}
\usepackage{bm}
\usepackage{epsf}
\usepackage{rotating}
\usepackage{epsfig,graphics,rotate,color}
\usepackage{wrapfig}
\usepackage{amssymb}
\usepackage{amsmath}
\usepackage{amsfonts}
\usepackage{array,hhline,dcolumn}%
\providecommand{\U}[1]{\protect\rule{.1in}{.1in}}
\begin{document}
\title{Reactor Simulation for Antineutrino Experiments using DRAGON and MURE}
\author{C.L. Jones$^1$, A. Bernstein$^2$, J.M. Conrad$^1$, Z. Djurcic$^3$, M. Fallot$^4$,
  L. Giot$^4$, G. Keefer$^2$, A. Onillon$^4$, L. Winslow$^1$}
\affiliation{$^1$Massachusetts Institute of Technology, 77 Massachusetts Avenue, Cambridge MA, 02139}
\affiliation{$^2$Lawrence Livermore National Laboratory, 7000 East Avenue, Livermore CA, 94550}
\affiliation{$^3$Argonne National Laboratory, 9700 S. Cass Avenue, Argonne IL, 60439}
\affiliation{$^4$Laboratoire SUBATECH, \'Ecole des Mines de Nantes, Universit\'e de Nantes, CNRS/IN2P3, 4 rue Alfred Kastler, 44307 Nantes
Cedex 3, France }


\begin{abstract}
Rising interest in nuclear reactors as a source of antineutrinos for experiments motivates validated, fast, and accessible simulations to predict reactor fission rates.  Here we present results from the DRAGON and MURE simulation codes and compare them to other industry standards for reactor core modeling.  We use published data from the Takahama-3 reactor to evaluate the quality of these simulations against the independently measured fuel isotopic composition. The propagation of the uncertainty in the reactor operating parameters to the resulting antineutrino flux predictions is also discussed.
\end{abstract}

\maketitle



As new high-power reactors come online, opportunities for
reactor-based antineutrino experiments are rising.  Three
experiments searching for the last unknown neutrino oscillation parameter,
$\theta_{13}$~\cite{PDG2010}, have released results \cite{DC2006, DC2011, DB2007, DBresult, RENO2010, renoresult}.
New short-baseline reactor oscillation
experiments~\cite{scraam} are motivated by the ``reactor antineutrino anomaly'', a recent analysis with results that are consistent with neutrino oscillations at $\Delta m^2\sim1$~eV$^2$~\cite{Anomoly2011}.  Searches for neutrino-nucleus coherent scattering~\cite{CoherentScatter2004} and studies of antineutrino-electron scattering~\cite{MUNU2004} using reactor sources are also underway.  
Precise measurements of antineutrino rates may also permit a real-time,
non-intrusive assay of the entire reactor core for nonproliferation
applications~\cite{reactorOrg,LLNL2009}.  

In the reactor core, neutron-rich fission products $\beta$-decay creating antineutrinos. The prediction of the antineutrino flux proceeds in two steps.
First, the fission rates of the primary fissile isotopes are calculated.  Then, this output is convolved with the
antineutrino spectrum, the sum of the spectra from the $\beta$-decay of each isotope's fission products.  The antineutrino spectral
predictions have recently been updated to include more detailed information on the daughter $\beta$-decay isotopes and higher-order corrections to the $\beta$ energy spectrum~\cite{Mueller2011,HuberReactor}.  In this paper, we focus on understanding the systematic uncertainties involved in the first step, the 
fission rate simulations. We introduce two codes: DRAGON~\cite{DRAGON1994}, a fast 2D parameterized simulation, and
MURE (MCNP Utility for Reactor Evolution)~\cite{Meplan2005,MURECode} a
3D Monte Carlo simulation.   While neutrino experiments require fission rate predictions, reactor core simulations  in industry 
focus on other quantities. In particular, the DRAGON code was modified by the authors to produce fission rates, whereas MURE already possessed this ability.  DRAGON and MURE are used in the recent Double Chooz result~\cite{DC2011}, and DRAGON is used by the Daya Bay experiment~\cite{DBresult}.

In this work, we compare our DRAGON and MURE simulations to the Takahama-3 benchmark.  
This benchmark allows a comparison of  absolute predictions of fissile material 
production to measurements from destructive assays of fuel rods 
from the Takahama-3 reactor in Japan~\cite{Takahama2001}. The Takahama-3 benchmark is the most complete and therefore most common data set to benchmark codes against, though other data sets exist~\cite{ArianeRebus}. By focussing on this benchmark,  we compare our results to those from proprietary reactor simulations used by industry, and demonstrate the quality of our predictions. This is an important step towards demonstrating that the predicted antineutrino fluxes are accurate.

\section{Overview of Fissile Isotope Production}
Oscillation experiments detect antineutrinos via
the signal: $\bar \nu_e + p \rightarrow e^+ + n$, which has a
threshold at 1.8~MeV.  Reactors produce antineutrinos above this threshold primarily through the decay chains of four isotopes: $^{235}$U, $^{238}$U, $^{239}$Pu and $^{241}$Pu.  However, we point out that both DRAGON and MURE are capable of simulating the full complement of fission products
produced during the evolution of a reactor core.  These include, but are not limited to, the long-lived isotopes:  $^{238}$Pu, $^{240}$Pu, $^{242}$Pu, 
$^{237}$Np, $^{239}$Np, $^{241}$Am,  $^{242}$Am, $^{242}$Cm and $^{243}$Cm as well
as the relatively short-lived uranium isotope $^{236}$U. 

The total fission rate of the four isotopes $^{235}$U, $^{238}$U, $^{239}$Pu and $^{241}$Pu is directly correlated with the total thermal power of the reactor. The exact fuel inventory has little effect on the total thermal power because the energies released per fission are very similar, 202.8~MeV per fission for $^{235}$U to 211.0~MeV per fission for $^{241}$Pu~\cite{energyPerFiss}. The antineutrino spectra per fission from these isotopes are significantly different.  Consequently, the detected spectrum of antineutrinos is affected by the fuel inventory. 

Most high-power reactors, including Takahama and the two Chooz reactors, are pressurized water
reactors (PWRs).   A PWR core is composed of approximately 200 assemblies, each assembly consisting of 
several hundred fuel rods.   Fresh fuel rods are typically composed of UO$_2$. All simulations, including DRAGON and MURE, 
require a specification of the initial fuel compositions and the
arrangement of the fuel rods within the assembly.   
Each assembly may also contain some number of instrumentation and control rods
for monitoring and controlling the conditions within the assembly.  The details of the assembly geometry are integral to a particular reactor design, and therefore are often characterized by the reactor manufacturer, for instance Westinghouse or Areva.

PWR fuel rods are constructed of cylindrical fuel pellets approximately 1~cm in diameter and 1~cm in length. Pellets are then stacked in the fuel rods. The structure of the rod is formed by the Zircaloy cladding. Zircaloy, chosen for its high melting point and transparency to neutrons, is composed of zirconium and trace amounts of chromium and tin. A fresh UO$_2$ pellet in a PWR typically
consists of uranium enriched to between 2.5\% to 4\% $^{235}$U by weight.  Specifying the fuel density is important to the simulation as it sets the total amount of fuel in the volume fixed by the cladding dimensions. The density of UO$_2$ is 10.96 g/cm$^3$ at 273 K. Simulations often use an effective fuel density which accounts for the
gross details of the fuel pellet packing and geometry. This effective density is called the pellet stack density. The default value for this quantity is ``95\% theoretical density'', and values from 9.98 g/cm$^3$ to 10.7 g/cm$^3$ are typical~\cite{Scale2010}.  Because the geometry of the rod is fixed, the pellet stack density determines the total amount of fuel.

Most PWRs burn a mixture of fresh fuel assemblies and assemblies that
have been through one or two fuel cycles, where a fuel cycle typically lasts about one year.  The assemblies, at varying stages of evolution, are arranged to produce a precise power distribution across the core. Re-burning the 
assemblies maximizes the energy
that is extracted from the fuel.  


The assembly from the Takahama core that was used in the benchmark began as a fresh assembly and proceeded through
three fuel cycles. This long irradiation makes this benchmark ideal for studying cumulative systematic effects. Systematic uncertainties come from three sources: uncertainties in the reactor data, theoretical uncertainties in the nuclear cross sections, and numerical approximations and methods used by the different codes.  Among the inputs from the reactor data, we focus on the moderator temperature and fuel density of the rod, and reactor core properties such as the power and moderator boron content.

\section{The Takahama Benchmark}
\begin{figure}
\begin{center}
\includegraphics[width=80mm]{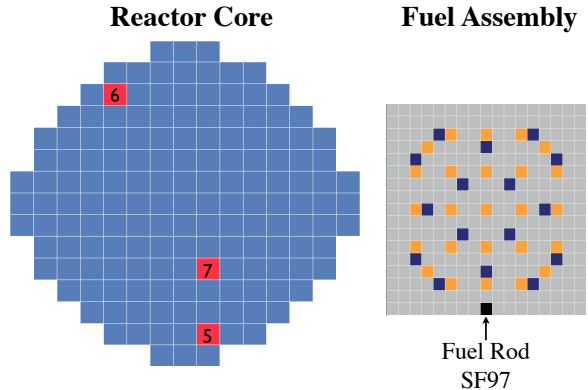}
\caption{\label{fig:TakahamaGeometry} Cross section of the Takahama reactor core with 157 fuel assemblies (left) and cross section of the fuel assembly (right). The location of the fuel assembly under study in fuel cycles 5, 6, 7 is indicated. In the fuel assembly, the position of the fuel rod referred to as `SF97' is shown in black, UO$_{2}$-Gd$_{2}$O$_{3}$ rods are shown in blue, and instrumentation rods are shown in orange.}
\end{center}
\end{figure}

The Takahama-3 reactor is a PWR that operates with 157 fuel
assemblies producing a total thermal power of 2652 MW. The assemblies
have a 17$\times$17 design, meaning there are 17$\times$17  locations
for rods. Diagrams of the Takahama core and of an assembly are shown in
Fig.~\ref{fig:TakahamaGeometry}. The benchmark began with assemblies
loaded with fresh UO$_2$ fuel rods with an initial
enrichment of 4.11\% $^{235}$U by weight, with the remainder being $^{238}$U with traces of $^{234}$U. Each assembly features 16
gadolinium-bearing (Gd$_2$O$_3$) fuel rods containing 2.6\%
$^{235}$U and 6\% gadolinium by weight.  The original publication shows 14 Gd rods~\cite{Takahama2001}; however, this number was updated to 16 in later publications~\cite{sfcomporef,sfcompo, sfcompocom}. 

Samples were taken from three fuel rods. Each sample was a 0.5~mm-thick disk.  Each sample was dissolved, and chemical separation was performed to isolate the isotopes of interest. Isotopic dilution mass spectroscopy was used to determine uranium and plutonium inventories; different mass spectroscopy and alpha and gamma counting techniques were used to determine isotopic concentrations of the other elements. For the most relevant isotopes, namely $^{235}$U, $^{238}$U, $^{239}$Pu, and $^{241}$Pu, the uncertainty associated with the determination of the isotopic mass fractions is $<$0.1\% for uranium isotopes and $< $0.3\% for plutonium isotopes~\cite{Takahama2001}.

The three fuel rods came from two different assemblies. From the first assembly, labeled NT3G23, a normal
uranium dioxide fuel rod (SF95) and a gadolinium-bearing fuel rod (SF96) were
studied after two cycles.   From the second assembly, labeled NT3G24, a normal uranium dioxide fuel rod (SF97) was studied after three cycles. We concentrate on SF97 because it has the longest irradiation time and therefore any cumulative systematic effects will be maximized. The rod was present
in three consecutive fuel cycles of 385, 402, and 406 days with 88
days and 62 days of cool-down time between cycles.  The location of SF97 within fuel assembly NT3G24 is shown in Fig.~\ref{fig:TakahamaGeometry} as is the location of fuel assembly NT3G24 in the three fuel cycles 5, 6, and 7. Samples were taken from SF97 at the six locations indicated in Table~\ref{tab:zaxismod}. Sample SF97-1 was located only 163~mm from the top of the rod, making the correct modeling of neutron leakage difficult. 

\begin{table}
\caption{\label{tab:zaxismod}Position of samples within the SF97 rod and the corresponding moderator temperature and burnup for that sample. Measurements are in
  mm from the top of the rod. The bottom of the rod is at 3863 mm.  The moderator temperatures are those for a theoretical light water reactor ~\cite{Takahama2001}. }
\begin{center}
\begin{small}
\begin{tabular}{l l l l l l l l}
\hline
Sample  &  Position & Mod. Temp. & Burnup \\
 & [mm] & [K] & [GW-days/ton] \\
\hline
1 & 163 & 593.1 & 17.69 \\
2 & 350 & 592.8 &  30.73 \\
3 & 627 & 591.5 &  42.16\\
4 & 1839 & 575.8 & 47.03 \\
5 & 2926 & 559.1 & 47.25\\
6 & 3556 & 554.2 & 40.79 \\
\hline
\end{tabular}
\end{small}
\end{center}
\end{table}

\begin{figure*}
\begin{center}
\includegraphics[width=160mm]{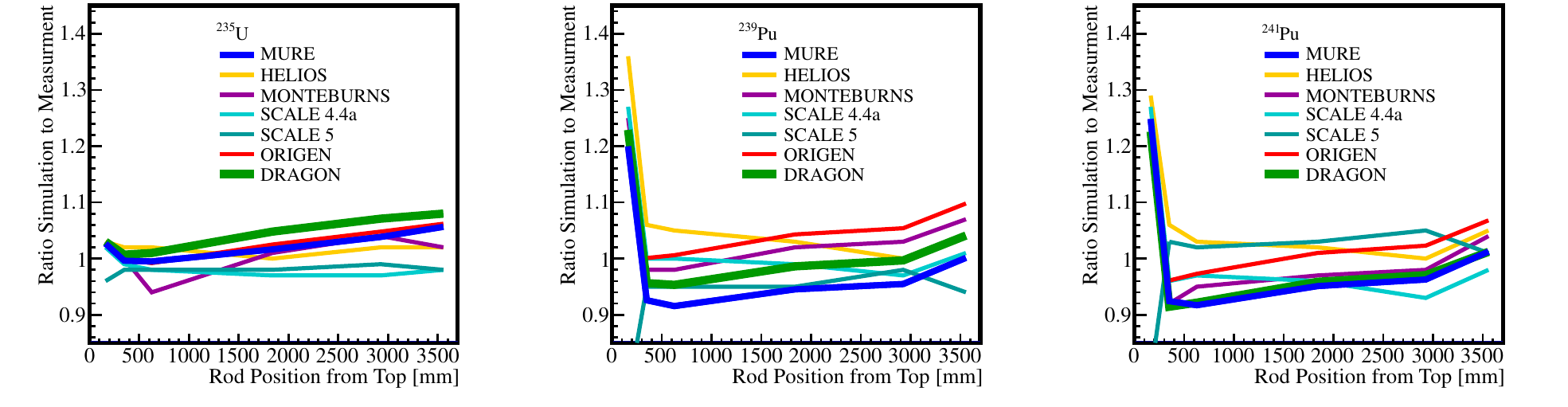}
\caption{\label{fig:CompareCode} Comparison of the ratio of calculated
  to measured mass inventories for SF97 for three isotopes important to antineutrino experiments: $^{235}$U, $^{239}$Pu and $^{241}$Pu.  DRAGON results are in green, and the
MURE results are in blue. Results from other published codes are overlaid for comparison. A linear interpolation between the six samples is used.}
\end{center}
\end{figure*}

\begin{table}
\caption{\label{tab:inputs} Takahama assembly parameters used as
  primary inputs to the DRAGON and MURE simulations.}
\begin{center}
\begin{small}
\begin{tabular}{l l}
\hline
Parameter & Value \\ 
\hline
Moderator Density & 0.72 g/cm$^3$ \\
Moderator Temperature & 600.0 K \\ 
Cladding Temperature &	600.0 K \\ 
Fuel Temperature &	900.0 K \\ 
Fuel Density &	10.07 g/cm$^3$ \\ 
Fuel Cell Mesh & 1.265 cm \\ 
Fuel Rod Radius & 0.4025 cm \\ 
Fuel Cladding Radius & 0.475 cm \\ 
Guide Tube Inner Radius & 0.573 cm \\ 
Guide Tube Outer Radius & 0.613 cm \\ 
Mean Boron Concentration & 630.0 ppm \\
\hline
\end{tabular}
\end{small}
\end{center}
\end{table}
The construction of the SF97 rod simulation starts with a
geometric description of the fuel assembly and the initial
isotopic inventory of the fuel pellets. The primary inputs used in the
simulations are found in Table~\ref{tab:inputs}. The power
history for each sample was determined via the $^{148}$Nd method~\cite{Takahama2001, TakahamaHeliosScale2001}.
This technique provides a detailed power history in time and along the length of the rod. The integrated exposure, or burnup, from this technique is summarized in Table~\ref{tab:zaxismod}.  We use a mean boron concentration
of 630 ppm per cycle~\cite{Takahama2001}. This is the standard value used by the other simulations 
considered in Sec.~\ref{compare}. For the pellet stack density we use 10.07
g/cm$^{3}$, 91\% of the theoretical density,  as suggested by Ref.~\cite{Roque2004}. This is lower than the standard 95\% of 10.96~g/cm$^3$, but is reasonable since the original paper Ref.~\cite{Takahama2001} does not specify the exact value. 

\section{Comparison of Reactor Core Simulation Codes\label{compare}}
Most deterministic codes, including DRAGON, simulate assemblies via a lattice calculation.  In a lattice calculation, one chooses a component, which is typically either a fuel rod or a fuel assembly.  The lattice component is assumed to give rise to a typical neutron flux, and therefore all surrounding components are identical, creating a lattice of these units.  A Monte Carlo code, like MURE,  simulates the neutron flux by actually generating and tracking neutrons.

Simulations are characterized by the number of dimensions used in the neutron transport equation they solve.  Thus, codes can be 1D, 2D (like DRAGON) or 3D (like MURE). A 1D simulation models the assembly with an effective lattice component 
rather than taking into account the actual shape.
A 2D simulation models a heterogeneous assembly,  taking
into account the cross sectional arrangement of the fuel
cells as is illustrated in Fig.~\ref{fig:TakahamaGeometry}.

The codes used for comparison in this study are SCALE 4.4a~\cite{ScaleHelios1998}, SCALE 5~\cite{Scale2010}, ORIGEN 2.1~\cite{ORIGEN},  MONTEBURNS ~\cite{Monteburns2009}, and HELIOS
~\cite{ScaleHelios1998}.  SCALE 4.4a is a 1D code with a detailed model of the
water/fuel geometry. It contains a separate module, SAS2H, for performing the fuel depletion calculations. SCALE 5 uses the 2D neutron transport model TRITON and the fuel depletion model NEWT.  Its validation is described elsewhere \cite{ScaleValidate, ScaleHelios1998}.  ORIGEN 2.1 is a fuel depletion code that models the buildup and decay of radioactive materials.  As such,  it does not model the neutron flux nor does it take the assembly geometry into account.  MONTEBURNS, like MURE, is a Monte Carlo code that uses MCNP-based transport.  It connects the transport abilities of MCNP with the depletion code ORIGEN 2.1. HELIOS version 1.6, like DRAGON, performs lattice calculations in a 2D plane and has a parameterized treatment of neutron transport. 

Many cross section libraries are available, including 
ENDF/B-VI~\cite{ENDFVI} and JENDL 3.2~\cite{JENDL},
and there is no consensus on the best choice.
The codes listed above use the following 
cross section databases as inputs: SCALE 4.4a and SCALE 5 use an ENDF/B-V library~\cite{ENDFV}. 
ORIGEN 2.1 uses JENDL 3.2, and MONTEBURNS and HELIOS 1.6 uses ENDF/B-VI-based libraries.    We discuss systematic uncertainties in final isotopic abundance due to the choice of cross section libraries in Sec.~\ref{syserr}.

\section{Comparison of Sample SF97 Simulation Results}
The DRAGON simulation for rod SF97 proceeded as follows.  The
simulation inputs were separated into two components: time-independent, such as the initial fuel loading and the pellet radii, and
time-dependent, such as the power and irradiation period.
The boron concentration was kept fixed at a non-burnable value of 630
ppm, the fuel temperature was kept at 900~K, and the moderator temperature was fixed at 600~K for all samples.  The input nuclear cross sections used in this simulation were ENDF/B-VI
with a WIMS-style~\cite{WIMS} transport correction, which accounts for the anisotropy in the scattering cross section in the laboratory frame.  The effect on the comparison to data is less than 1.5\% for all isotopes. A correction for molecular effects is also included. Results using
JENDL 3.2~\cite{JENDL} cross section libraries are shown for comparison.  

The simulation reads in the power in time steps provided in Ref.~\cite{Takahama2001}.
The status of the simulation for each step is saved and used as the input to the next step.  Only the final step of the simulation can be compared against
the destructive assay data. However, we can use the results of all intermediate
evolution steps to evaluate systematic effects in the fission rate studies as presented below.

By exploiting the symmetry of the assembly, we can model an $1/8$ segment and save computation time. In each step, the neutron flux in the segment is evolved
using the collision probability method with self-shielding
corrections~\cite{DRAGON1994}.  After the flux is computed, the fuel depletion module
evolves the isotopic composition of the fuel by solving the Bateman equations using a Runge-Kutta method.  The calculation for the full three fuel cycle evolution takes 27.5 hours on a 2.8-GHz processor. At this time, the DRAGON simulation has not been parallelized.

The MURE simulation proceeds similarly. Instead of an $1/8$ segment, the full assembly is simulated in 3 dimensions with specular boundary conditions on all surfaces of the assembly. The height of the assembly was taken to be 1~cm and a different simulation was run for each sample.  This effective 2D model is used to allow a comparison between deterministic versus Monte Carlo approaches. The MURE simulation starts with the generation of $10^{5}$ neutrons.  Using MCNP, these neutrons are tracked from the parent fission process until they are absorbed.  This cycle of neutron generation and tracking is repeated 1900 times to ensure an equilibrium state is reached. At this point, an additional 100 cycles using $10^6$ neutrons are used to calculate the parameters of interest for this time step. The fuel evolution is then calculated by solving the Bateman equations using a Runge-Kutta method. The input nuclear cross sections are once again ENDF/B-VI with molecular effects. Simulations with JENDL 3.2~\cite{JENDL} are shown for comparison. Though MURE can use continuous cross sections, a multi-group treatment is used to increase the speed of these simulations. It uses 179,000 neutron groups (in comparison to DRAGON's 172 groups). The effect of the multi-group treatment compared to running with continuous cross sections is negligible. For the sensitivity studies in Section~\ref{syserr}, the number of neutrons is reduced to $10^4$ and results are averaged over the assembly. The reduction in simulated neutrons increases speed, and the full three cycle evolution takes 9 hours using 10 2.5-GHz processors. 

When the MURE and DRAGON simulations are complete, the results for rod SF97 are extracted. Fig.~\ref{fig:CompareCode} shows the ratio of calculated to
experimentally-measured mass inventories. The results for $^{238}$U are not shown since its mass does not deplete by more than 0.1\%. This is of the same order as the uncertainty in the mass inventory, and therefore does not yield a useful comparison. For the other isotopes, the DRAGON and MURE results are consistent with 
the data along the rod. However, there is a large deviation in SF97-1, located near  the top edge of the fuel rod, which arises from approximations of neutron leakage in the axial dimension. This effect is observed in results from all the codes. As is discussed in Section~\ref{sec:fiss}, the contribution of SF97-1 to the number of fissions is only a third of that from the other five samples, and the increased uncertainty is negligible.

Neglecting SF97-1, we calculate the average deviation over the rod by taking the average of the samples. For $^{235}$U, the codes range from -2.2\% to 4.5\% with MURE at  2.1\% and DRAGON at 4.3\%. Even neglecting sample 1, deviations for $^{239}$Pu  range from MURE at -5.1\% up to 6.5\% for ORIGEN, while DRAGON has a deviation of -1.3\%. Finally, for $^{241}$Pu the codes range from -4.6\% up to 3.4\% with MURE and DRAGON at -4.6\% and -4.4\% respectively.  

Since a principal aim of this work is the prediction of quantities useful to reactor antineutrino experiments, we have ensured that the simulation inputs are identical between DRAGON and MURE. The libraries used by the Monte Carlo codes only contained moderator cross section information evaluated at 600~K.  Thus, in order to maintain identical inputs between our simulations, the DRAGON simulation used a moderator temperature of 600~K as well for all six samples.  In fact, the moderator temperature varies along the rod as shown in Table~\ref{tab:zaxismod}.  The SCALE simulations used the more detailed moderator temperature and calculated the corresponding moderator density change~\cite{TakahamaHeliosScale2001}. This may explain the better performance of this code.


\section{Systematic Uncertainties\label{syserr}}
\begin{figure*}
\begin{center}
\includegraphics[width=160mm]{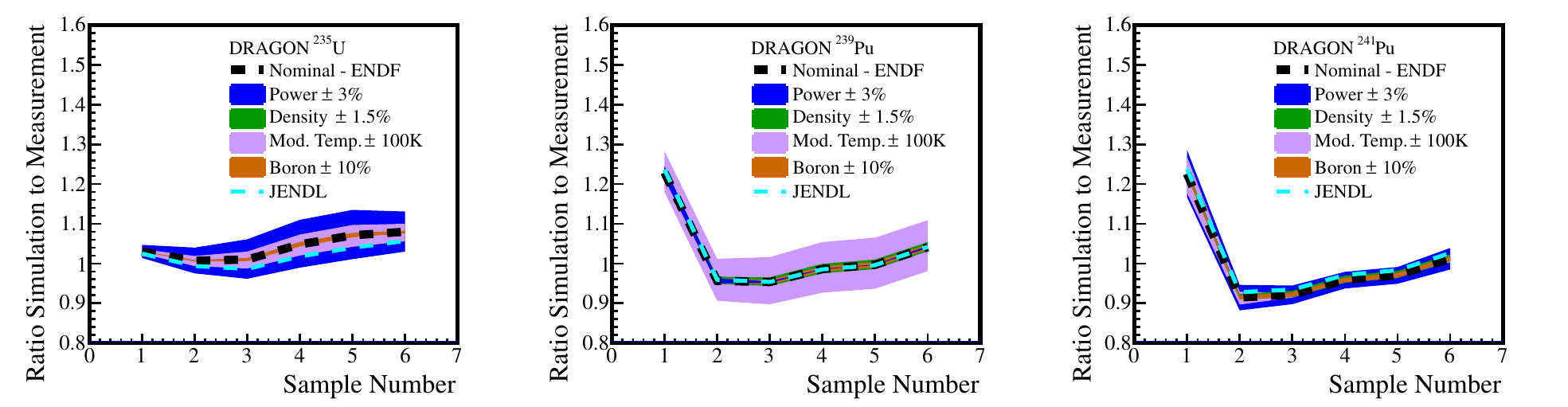}
\includegraphics[width=160mm]{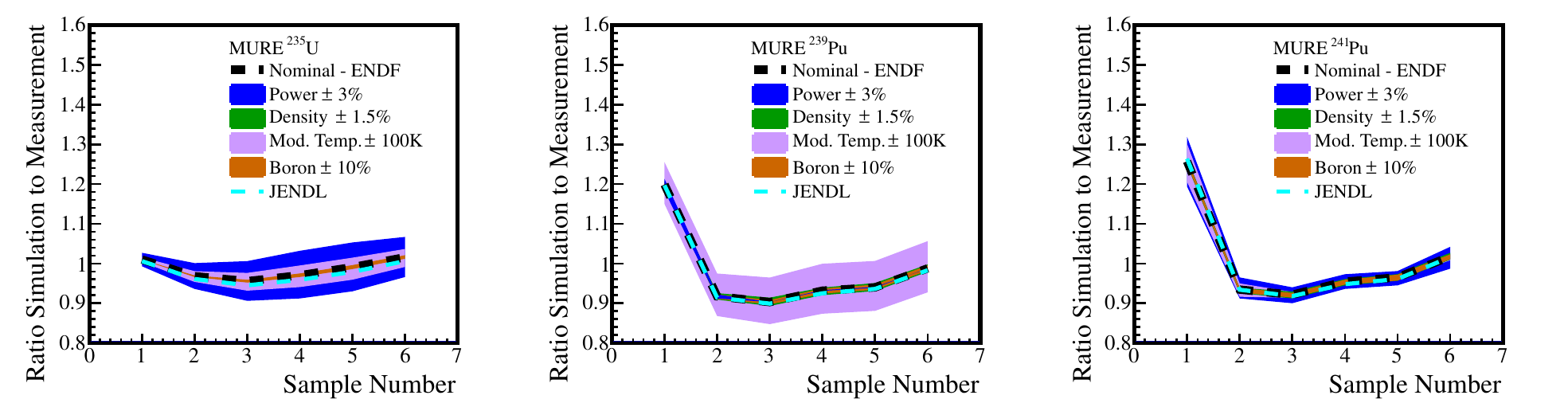}
\caption{\label{fig:SenseReact} Sensitivity of the Takahama benchmark
  predictions for SF97 to four uncertainties on the fuel rod design
  and operation. The uncertainties are overlaid. MURE results use the average over the full assembly. The sensitivity is plotted as a function of sample number, which can be a proxy for both the axial position along the rod as well as the burnup reached in that sample.}
\end{center}
\end{figure*}

\begin{table}
\caption{\label{tab:SenseReact}Study of the systematic effect of varying the thermal power, fuel density, moderator temperature and boron concentration on the mass inventory for SF97-4. The ratios of the varied simulation to the nominal simulation are shown. MURE results use the average over the full assembly.}
\begin{center}
\begin{small}
\begin{tabular}{l l l l l l l}
\hline
& \multicolumn{2}{ c }{ $^{235}$U} &\multicolumn{2}{ c }{$^{239}$Pu} & \multicolumn{2}{ c }{$^{241}$Pu} \\
  			&{\scriptsize MURE}&{\scriptsize DRAGON}&{\scriptsize MURE}&{\scriptsize DRAGON}& {\scriptsize MURE}&{\scriptsize DRAGON}\\ 
\hline
\multicolumn{7}{ c }{ Thermal Power } \\
\hline
$+3\%$& 0.940 & 0.944 & 0.999& 1.001 & 1.020 & 1.021 \\
$-3\%$& 1.063 & 1.059 & 1.001 & 0.999 & 0.981 & 0.978 \\
\hline
\multicolumn{7}{ c }{ Fuel Density } \\
\hline
$+1.5\%$& 0.992 & 0.991& 0.988 & 0.989 & 0.990 & 0.992\\
$-1.5\%$& 1.007 & 1.009 & 1.008 & 1.013 & 1.007 & 1.011\\
\hline
\multicolumn{7}{ c }{ Moderator Temperature } \\
\hline
$+100$K& 1.025 & 1.024 & 0.938 & 0.940 & 1.000 & 1.002 \\
$-100$K& 0.969 & 0.973 & 1.073 & 1.069 & 0.994 & 0.996 \\
\hline
\multicolumn{7}{ c }{ Boron Concentration } \\
\hline
$+10\%$& 1.005 & 1.006 & 1.006 & 1.007& 1.005 & 1.006\\
$-10\%$& 0.995 & 0.995 & 0.992 & 0.993 & 0.993 & 0.994\\
\hline
\end{tabular}
\end{small}
\end{center}
\end{table}

The primary systematic uncertainties in describing a reactor core for simulating
fission rates were identified in Ref.~\cite{Zelimir2009}.  We have
reconfirmed that these are the major uncertainties in describing the core and present specifics for the Takahama benchmark. Beyond this, we present a systematic study of the effect of 
cross section uncertainties on the result.

The most significant uncertainties related to the reactor
core description arise from the thermal power and the temperature 
of the moderator. The density of the fuel and the mean boron loading are secondary effects.  The uncertainty in the specific thermal power is taken to be 3\%, the uncertainty of the $^{148}$Nd method~\cite{Nd148uncert}.  Usually, simulations are given individual assembly power densities and the full core thermal power.  The uncertainty for the thermal power for the core is typically $<$2\%~\cite{Zelimir2009}.  Though the $^{148}$Nd method has a larger uncertainty, it allows for a more detailed study of the response along the rod and is the only power information available for the benchmark. The variation of moderator temperature along rod SF97 is approximately 50 K~\cite{Takahama2001}.  However, the continuous cross section libraries only contained evaluated data at 500~K and 700~K, and so we performed a $\pm100$ K variation for the moderator with both codes. MURE used ENDF/B-VII~\cite{ENDFVII} libraries that included molecular effects to improve the water model in this study and ENDF/B-VI for the other nuclei. The density of the moderator was kept constant.

We have chosen to vary the fuel density by 1.5\%~\cite{Density}. Both DRAGON and MURE contain parameters that can explicitly vary the fuel density.  However, a simple variation of this density parameter changes the total amount of fuel.  Since we seek to compare our results against an empirical determination of the inventory, we have instead elected to vary the fuel density and fuel rod radius simultaneously, while keeping the initial mass of uranium constant. The boron variation used for the study is 10\%~\cite{Takahama2001}. 

In Table~\ref{tab:SenseReact}, the results are summarized for sample SF97-4. MURE ran with smaller statistics and averaged over the all of the rods of the assembly. Power variation is particularly important to the uncertainty on $^{235}$U.  This follows from the fact that $^{235}$U is the primary reactor fuel and drives the thermal power. Since $^{239}$Pu and $^{241}$Pu are the products of neutron reactions on $^{238}$U, they are more sensitive to changes in temperature.   Increasing the amount of boron in the moderator will prohibit thermal fissions.  The 10\% variation in the mean boron concentration leads to a small effect, less than 0.5\%.    

Fig.~\ref{fig:SenseReact} shows the effect of varying the above
inputs within the systematic errors on the prediction for the 
SF97 data.  
The variations are overlaid, with the smallest effects shown on top of larger variations.  In the case of $^{235}$U, the variations contain both the measured mass inventory and the spread in other codes. For $^{239}$Pu, the larger 100 K variation in temperature would include both measurements and the results of the other codes. In addition, the total mass of uranium can be increased by 4\% by increasing the pellet stack density, improving the agreement for both plutonium isotopes.

To examine the effect of moderator temperature on our benchmark results, we held the moderator density fixed.  The effect is large for $^{239}$Pu because changing the moderator temperature affects the neutron capture cross section of $^{238}$U, which drives the production of $^{239}$Pu.  In contrast, $^{235}$U is the primary source of fissions, and so it is more sensitive to a power variation.  We find that for $^{235}$U, the uncertainty always grows.  For power variations, we see the uncertainty in the plutonium isotopes reduces along the rod axis. We note that in the cases where the systematic uncertainties are large, the masses are smaller and therefore the average effect is small.  

We have also examined uncertainties arising from the fission and capture
cross section inputs.  As mentioned in Sec.~\ref{compare}, there is
no consensus on the best choice of library.    However, we note that
all libraries are evaluating the cross sections based on the same data sets, and so 
are highly correlated. In Fig.~\ref{fig:SenseReact}, we compare the nominal ENDF/B-VI~\cite{ENDFVI} to JENDL 3.2~\cite{JENDL}.  The difference between cross section libraries is most important for $^{235}$U, causing a 1.1\% change in sample SF97-4 for MURE and a 3.0\% effect for DRAGON. For $^{239}$Pu, DRAGON shows a 0.1\% effect while MURE sees a 0.7\% effect. Finally, for $^{241}$Pu, MURE and DRAGON see a 0.6\% and 1.2\% effect respectively. 

MURE can easily modify the energy released per fission which is used to tie the fission rate to the thermal power measurement.   In DRAGON, this is more difficult as it is integrated into the calculation with a particular cross-section library. To understand the effect of these values, MURE was run with energies per fission as calculated by DRAGON, and found this to be a 1\% effect for SF97-4.

\section{Correspondence Between Mass Inventory and Fission Rates\label{sec:fiss}}
\begin{figure*}
\begin{center}
\includegraphics[width=160mm]{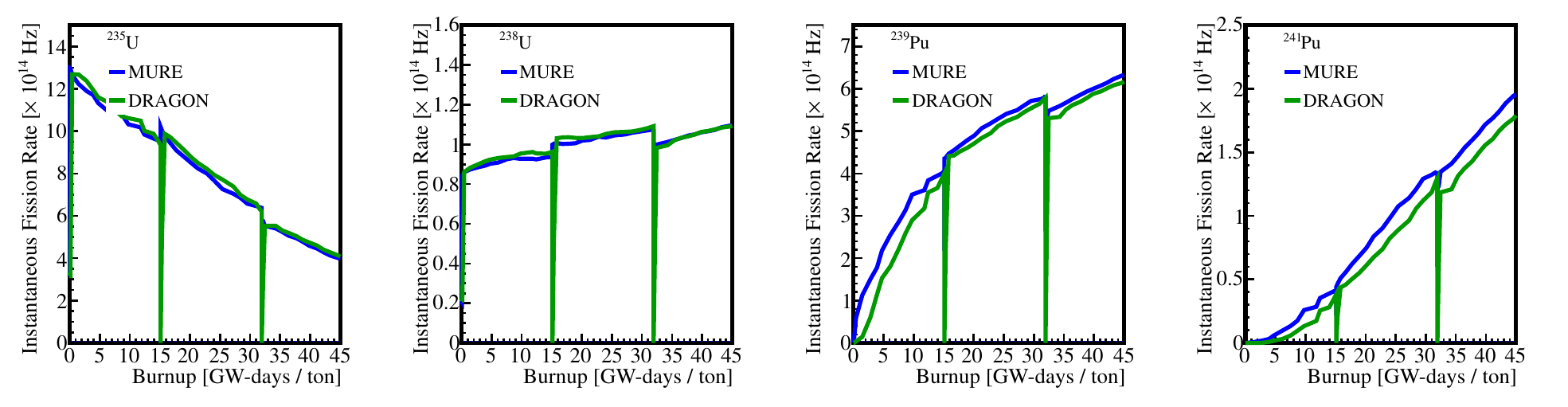}
\includegraphics[width=160mm]{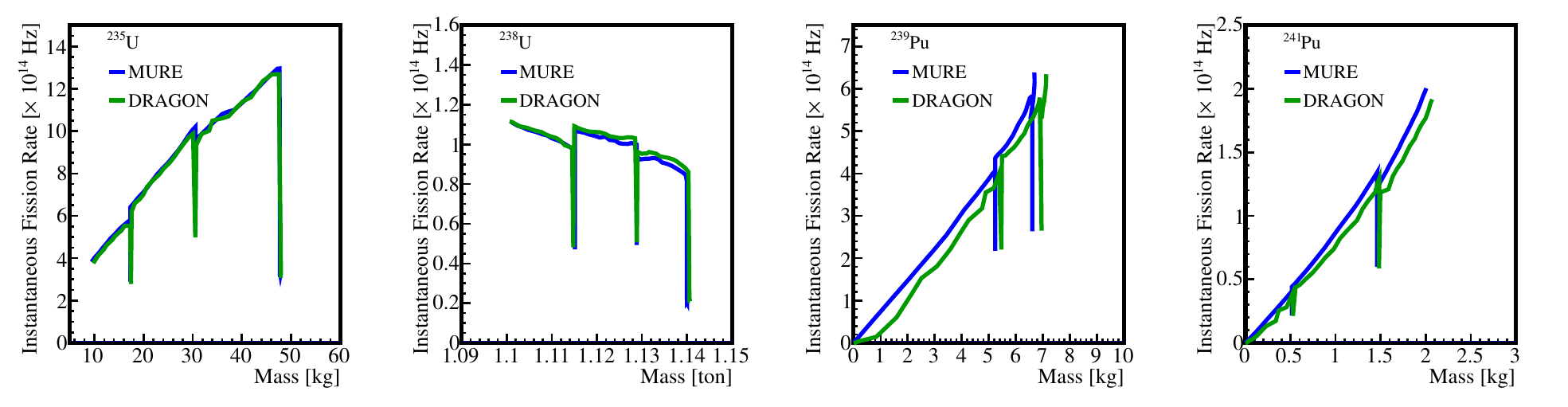}
\includegraphics[width=160mm]{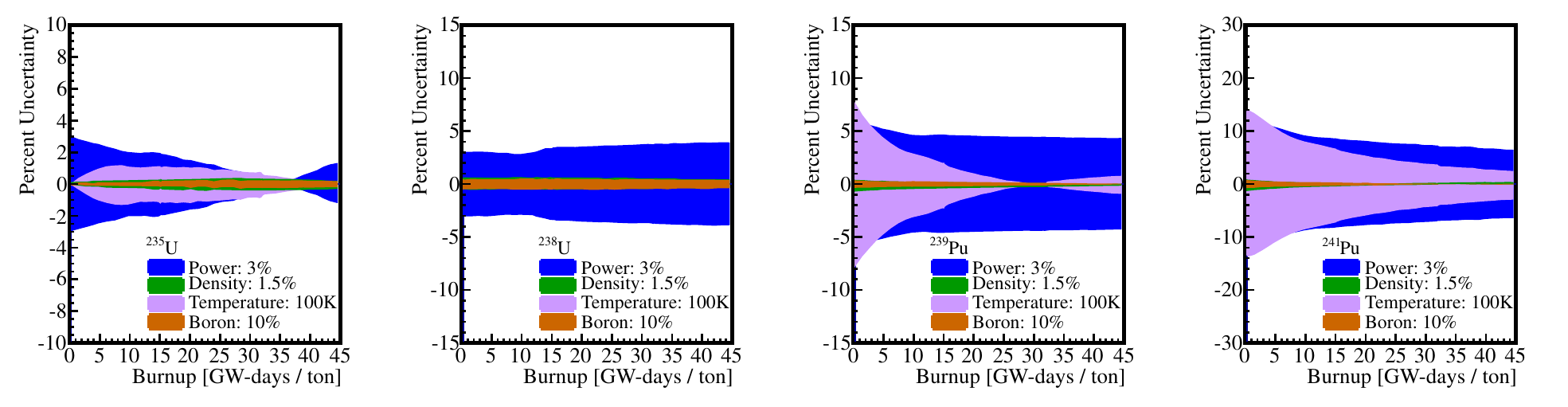}
\caption{\label{fig:FissionVsMass} Assembly-level instantaneous fission rates from DRAGON and MURE for the Takahama simulation for SF97-4. Top:  Instantaneous fission rates shown for the nominal simulation.  Middle: The correlation between the instantaneous fission rate and resulting mass is shown. Bottom: Sensitivity of the instantaneous fission rates to the major uncertainties in the simulation inputs. The DRAGON results are shown and are consistent with MURE.}
\end{center}
\end{figure*}

This paper has used comparisons of the DRAGON and MURE predictions
of mass inventories to the Takahama data to demonstrate the quality
of the simulations. However, neutrino experimenters are interested 
in fission rates rather than mass inventories. In the upper portion of Fig.~\ref{fig:FissionVsMass}, we show the instantaneous fission rates as a function of burnup for the assembly containing SF97.  This simulation used the inputs for SF97-4 through the three fuel cycles. For a given fuel assembly, the fissions of $^{235}$U dominate the antineutrino production until the beginning of the third fuel cycle, when the $^{239}$Pu contributes equally. This occurs at a burnup of $\sim$35~	GW-days / ton.  The fissions from $^{238}$U and $^{241}$Pu contribute approximately 10\% of the flux until the end of the third fuel cycle when they reach parity with those from $^{235}$U. 

The difference in instantaneous fission rates between MURE and DRAGON is on average 2.6\% for $^{235}$U, 2.9\% for $^{238}$U, 4.9\% for $^{239}$Pu and 9.5\% for $^{241}$Pu. The differences are largest during the first fuel cycle. Since great care was taken to use the same inputs, this can be used as a systematic uncertainty between Monte Carlo based codes like MURE and deterministic codes like DRAGON.  We note that with some tuning of the inputs and evolution step size, some reductions in this uncertainty are possible. 

Fundamentally, the Takahama benchmark is a test of a code's ability to simulate a fuel assembly. There is insufficient information about the fuel inventory and power distribution over the full core to make statements about systematic uncertainties across it. In general, fuel assemblies at the edge of the core have fission rates that are $\sim$50\% less than those at the center due to power variation across the core, and variations of $\sim$10\% are expected between neighboring assemblies due to fuel inventory differences.  If given this more detailed information, a full-core simulation can be constructed, as was done for~\cite{DC2011} and will be discussed in future work by the authors~\cite{FutureReactorPaper}.  It is also difficult to make statements about fuel rods other than SF97. The power input as a function of $z$ comes from the $^{148}$Nd method, and we have this information for only rod SF97. By construction, all rods in the assembly have the same power distribution, and the assembly-averaged integrated number of fissions is the same as those from rod SF97.  Since burnup is a proxy for the number of fissions, the distribution of fissions in z must also agree with the provided burnup values for the samples~\cite{Takahama2001}.  We see in Table~\ref{table:fissionZ} that the results of the simulation are consistent with the integrated burnup.

For the individual fuel rod SF97, we can make statements about the integrated number of fissions as a function of $z$. The axial component of the fission rate $F$ is proportional to the axial component of the neutron flux.  For an ideal cylindrical reactor, this axial component can be described analytically: $F \propto \sin \frac{\pi z}{H}$ \cite{AxialFlux} where $z=0$ is defined as the top of the core and $H = 403$~cm is the total height of the core. The assembly-averaged integrated number of fission from DRAGON, $\int F(t) \,dt$, is compared to the analytical calculation in Table~\ref{table:fissionZ}. The results across the rod are more flat for the simulation than for the analytic calculation. The contribution of SF97-1 to the total integrated number of fissions is less than half of the contribution from SF97-4, thus the larger uncertainties on this sample are mitigated by its lower contribution to the total antineutrino flux.  

\begin{table}
\caption{\label{table:fissionZ} The assembly-averaged total integrated number of fissions from DRAGON and the integrated burnup over the three fuel cycles from~\cite{Takahama2001} as a function of $z$ and normalized to sample SF97-4.  The final column shows the axial neutron flux along the $z$-axis, also normalized to SF97-4.}
\begin{tabular}{ccccc}
\hline
Sample & $z$ & Fissions & Burnup & $\sin (\pi z / H )$ \\ 
 & [cm] & [\%] & [\%] & [\%] \\
\hline
1 & 16.3 & 38.4 & 37.6 & 12.78 \\ 
2 & 35.0 & 66.0 & 65.3 & 27.17 \\ 
3 & 62.7 & 89.9 & 89.6 & 47.35 \\ 
4 & 183.9 & 100.0 & 100.0 & 100.00 \\ 
5 & 292.6 & 100.5 & 100.5 & 76.79 \\ 
6 & 355.6 & 87.0 & 86.7 & 36.86 \\ \hline
\end{tabular}
\end{table}

The correlation between the instantaneous fission rates and the mass inventories is what permits us to use the measured mass inventories to evaluate the performance of these codes.  The mass inventories and the instantaneous fission rates maintain a linear correlation to first order over the three fuel cycles. This is shown in the middle part of Fig.~\ref{fig:FissionVsMass}. It is this relationship that allows antineutrino detectors to monitor the mass inventories in reactors for non-proliferation applications. To understand the systematic uncertainties in the fission rates, we vary the input parameters as was done in Section~\ref{syserr} for the mass inventories. The results of this study at the end of three fuel cycles are summarized in Table~\ref{tab:SenseMureFiss}, and, as with the mass variation studies, the major systematic uncertainty is the thermal power. 

The systematic uncertainties are not constant as a function of burnup, as shown in the bottom part of Fig.~\ref{fig:FissionVsMass}.  This effect is also seen in the mass studies when comparing the samples with different burnup values.  During the first fuel cycle, the moderator temperature variation is the largest systematic uncertainty for the plutonium isotopes, but it is not a comparable effect for $^{235}$U until a burnup of 20~GW-days / ton, halfway through the second fuel cycle.  The sensitivity plots have an intersection when the upper and lower variations coincide.  This crossover occurs because all variations use the same initial amount of fuel and are simply evolving it at different rates according to the varied parameter.

The technique of varying the inputs of the simulation to determine the correlated uncertainty is applicable to all reactor antineutrino analyses. However, setting a systematic uncertainty on the fission rates from the benchmark is difficult since the mass inventories are only available at the end of three fuel cycles for a limited number of fuel rods.  Also, the Takahama benchmark has a 3\% uncertainty in the thermal power, which is determined from the $^{148}$Nd method.  This value is larger than the typical $\approx 0.7\%$ from standard reactor instrumentation~\cite{Zelimir2009}.  The benchmark also lacks detailed density information.  This leads to the large systematic uncertainties in the fission rates shown in Fig.~\ref{fig:FissionVsMass}.  For these reasons, the benchmark is used to understand the systematic uncertainty from using different codes, and to provide an upper limit on the systematic uncertainties for full-core simulations.

\begin{table*}
\caption{\label{tab:SenseMureFiss}Study of the systematic effect of varying the thermal power, fuel density, moderator temperature and boron concentration on the fission rates for SF97-4. The ratios of the varied simulation to the nominal simulation are shown. MURE results use the average over the full assembly.}
\begin{center}
\begin{small}
\begin{tabular}{lllllllll}
\hline
& \multicolumn{2}{ c }{ $^{235}$U} & \multicolumn{2}{ c }{ $^{238}$U} &\multicolumn{2}{ c }{$^{239}$Pu} & \multicolumn{2}{ c }{$^{241}$Pu} \\
  			&{\scriptsize MURE}&{\scriptsize DRAGON}&{\scriptsize MURE}&{\scriptsize DRAGON}& {\scriptsize MURE}&{\scriptsize DRAGON} & {\scriptsize MURE}&{\scriptsize DRAGON}\\ 

\hline
\multicolumn{9}{ c }{ Thermal Power } \\
\hline
$+3\%$& 0.981 & 0.987 & 1.038 & 1.039 & 1.043 & 1.044 & 1.065 & 1.065 \\
$-3\%$& 1.014 & 1.012 & 0.958 & 0.961 & 0.957 & 0.957 & 0.938 & 0.936 \\
\hline
\multicolumn{9}{ c }{ Fuel Density } \\
\hline
$+1.5\%$& 1.002 & 1.002 & 0.996 & 0.995 & 0.999 & 1.000& 1.002 & 1.004 \\
$-1.5\%$& 0.999 & 0.997 & 1.002 & 1.003 & 1.000 & 1.002 & 0.999 & 0.999 \\
\hline
\multicolumn{9}{ c }{ Moderator Temperature } \\
\hline
$+100$K& 1.005 & 1.001 & 1.002 & 1.001 & 0.990 & 0.992 & 1.021 & 1.025 \\
$-100$K& 0.998 & 0.999 & 1.001 & 0.994 & 1.012 & 1.009 & 0.970 & 0.971 \\
\hline
\multicolumn{9}{ c }{ Boron Concentration } \\
\hline
$+10\%$& 0.998 & 0.998 & 0.997 & 1.004 & 1.001 & 1.000 & 0.999 & 0.999 \\
$-10\%$& 1.002 & 1.002 & 0.995 & 0.996 & 1.000 & 1.000 & 0.996 & 1.001 \\
\hline
\end{tabular}
\end{small}
\end{center}
\end{table*}


\section{Conclusions}
This paper has demonstrated the quality of  
two codes available for use in the prediction of reactor antineutrino fluxes, MURE and
DRAGON. We have established that MURE and DRAGON make accurate predictions based on their comparison to the well-known
Takahama benchmark.   They reproduce the mass inventory of 
rod SF97 to the level of other widely-used codes.  We
have demonstrated how these 
codes can be used to study systematic errors associated with the
reactor flux predictions.    We have confirmed that the thermal power is
the dominant contributor to the overall uncertainty in the prediction
of the mass inventory for the Takahama assembly.  We have shown that the mass
inventory tracks the fission rates, and thus the thermal power uncertainty 
can be expected to be the most important issue in predicting the flux
for neutrino oscillation experiments.  We have ensured that the simulations use identical inputs, and have thus provided a study of the difference between deterministic and Monte Carlo codes.

This paper has demonstrated the high quality
of the simulations;  however, the results presented
in this paper are specific to the Takahama benchmark.
General conclusions about fission rates
and uncertainties cannot be drawn as each reactor
core and fuel cycle is unique.  Instead, we encourage
neutrino experimenters to acquire the DRAGON and
MURE codes to model their individual reactor cores.

\section*{Acknowledgements}
The authors thank the NSF, DOE, CNRS/IN2P3 and PACEN/GEDEPEON for their generous support.  The authors thank the Double Chooz collaboration for their valuable input. The authors also thank Dr. Guy Marleau and Dr. Ben Forget for extensive support in using the DRAGON code.

\bibliographystyle{apsrev}
\bibliography{TakahamaPaper_PRD}

\end{document}